\documentstyle[12pt]{article}
\begin{document}
\normalbaselineskip=16 true pt
\normalbaselines

\def\cone {\chi_1^0}
\def\ctwo {\chi_2^0}
\def\mc {m_{\chi_1^0}}
\def\L {\Lambda} 
\def\s {\sigma}
\def\grav {\tilde G}
\def\gtil {\tilde \Gamma}

\def\be {\begin{equation}}
\def\bea {\begin{eqnarray}}
\def\ee {\end{equation}}
\def\eea {\end{eqnarray}}

\def\b {\bibitem}

\def\r {\rightarrow}
\def\lr {\longrightarrow}


\begin{flushright}
MRI-PHY/P981067\\
JU/HEP-TH-98/2
\end{flushright}

\vskip 1 true cm

\begin{center}

{\Large\bf
{A GEOMETRICAL INTERPRETATION OF PARITY VIOLATION IN GRAVITY WITH
TORSION}}\\[10mm]
{\large\em  Biswarup Mukhopadhyaya$^{*}$\footnote{Electronic address: 
biswarup@mri.ernet.in} and Soumitra Sengupta$^{*}$\footnote{Electronic 
address: 
soumitra@juphys.ernet.in}}\\[5mm]
$^{*}$ Mehta Research Institute, Chhatnag Road, 
Jhusi, Allahabad - 221 019, India\\
$^{**}$ Physics Department, Jadavpur University, Calcutta - 700 032, India
\end{center}

\begin{abstract}
In a space-time with torsion, 
the action for the gravitational field can be extended with a 
parity-violating piece. We show how to obtain such a piece
from geometry itself, by suitably modifying the
affine connection so as to include a pseudo-tensorial part. A consistent 
method is thus suggested for incorporating parity-violation in the
Lagrangians of all matter fields with spin in a space-time background 
with torsion.

\end{abstract}

\vskip 1.0 cm
\begin{center}
\end{center}

\normalbaselineskip=18 true pt
\normalbaselines
\newpage
\setcounter{footnote}{0}

Theories of gravitation in a space-time with torsion have been under
investigation for a long time. The presence of torsion modifies the spacetime
geometry through non-symmetric connections\cite{asym}. Looking from 
another angle,  the 
commutator of the covariant derivatives acting on a scalar does not vanish in 
a torsioned space. Torsion has been incorporated in theories of gravitation, 
which range from ones as old as the Einstein-Cartan model to delvelopments 
as recent as superstring theories.

It has been argued that torsion is an inescapable consequence 
if the matter fields giving rise to space-time curvature are possessed with
spin\cite{spin}. Various consequences of 
torsion have already been explored in the literature, including those 
involving complex connections\cite{comp}, and 
aiming, for example, to present the 
electromagnetic potential as the trace of the torsion tensor\cite{em}.
One of the most important observations in this context is that
the presence of torsion destroys the cyclic
property of the Riemann-Christoffel tensor. As a result, the
standard Einstein-Hilbert action admits of an additional
term which is {\it parity-violating}\cite{pv}. 
Thus a generalised theory of gravity can be conceived of as having the
possibility of parity violation inbuilt in it. An immediate extension of this
idea leads to the expectation that the Lagrangians of various types 
of matter fields coupled to
a torsioned space-time background\cite{matter1,matter2} 
will also contain parity-violating terms.
Although such terms are {\it a priori} weaker than those corresponding to
weak interacton (which is an already established  source of parity violation),
they can be significant in astrophysical and  cosmological contexts\cite{apc}, 
particularly in relation 
to coherent effects. And, above all, the sheer universality of gravity 
compels us to attach sufficient importance to such new physics possibilities.

The question that perhaps may be raised is: is there a consistent way of 
introducing parity violation in a spacetime with torsion, which will 
be applicable to both pure gravity and the various types of matter 
fields with spin?
In this note, we try to answer this question by seeking the origin of 
parity (or any discrete symmetry) violation in the geometry of space-time. 
We point out that just as the original Einstein-Cartan theory follows by
adding an antisymmetric tensor to the hitherto symmetric affine connnections,
the incorporation of a pseudo-tensorial connection introduces parity violation
in the theory. This enables one to start from the original Einstein-Hilbert
action, with the scalar curvature $R$ suitably modified by the new form of
the covariant derivative. We show that, with the most general choice of the
pseudo-tensorial connection which is linear in the torsion field, the
induced parity-violating term has close resemblance to what arises
in earlier works from a separate piece in the Lagrangian. 
This generalised connection also enables us to obtain the Lagrangians for 
spin-1 and spin-1/2
fields, with built-in parity-violating interactions with the
torsion field. Thus one can trace  all parity-violating effects in torsioned 
space to a common origin.     
 
The Einstein-Hilbert action for pure gravity is given by

\begin{equation}
S = \int \sqrt{-g} R d^{4}x 
\end{equation}

\noindent
where $R$ is the scalar curvature, defined as 
$R = R_{\alpha\mu\beta\nu}g^{\alpha\beta}g^{\mu\nu}$. $R_{\alpha\mu\beta\nu}$ 
is the Riemann-Christoffel tensor:

\begin{equation}
R^{\kappa}_{\mu\nu\lambda} = \Gamma^{\kappa}_{{\nu\lambda},\mu}
- \Gamma^{\kappa}_{{\mu\lambda},\nu} + 
\Gamma^{\kappa}_{\mu\sigma} \Gamma^{\sigma}_{\nu\lambda} - 
 \Gamma^{\kappa}_{\nu\sigma} \Gamma^{\sigma}_{\mu\lambda}  
\end{equation}    

In the absence of torsion, the $\Gamma$'s are the usual Christoffel
symbols, symmetric in the two lower indices.

If there is torsion, $\Gamma$ needs to be replaced by
$\gtil$, where 
 
\begin{equation}
\gtil^{\mu}_{\nu\lambda} = \Gamma^{\mu}_{\nu\lambda} - H^{\mu}_{\nu\lambda}
\end{equation}

\noindent
$H$ being antisymmetric in the two lower indices. Further, the requirement
that the metric be covariantly conserved (the so-called `metricity condition')
restricts $H$ to a form where it is antisymmetric in all three indices, 
a form in which it is commonly known as the `contortion tensor'. It is 
related to the torsion field S 
through $H_{\mu\nu}^{\lambda} = -S_{\mu\nu}^{\lambda} +
{S_{\nu}}^{\lambda}_{\mu} - S^{\lambda}_{\mu\nu}$.

The inclusion of H destroys the cyclicity of $R_{\alpha\beta\mu\nu}$
in any three of its four indices. Consequently, a term of the form 
$\epsilon^{\alpha\beta\mu\nu} R_{\alpha\beta\mu\nu}$ (which, in a
torsion-free scenario, would have been 
forced to vanish by the cyclicity condition) can be added to the scalar
curvature \cite{pv} , in perfect consistence with general covariance.  
The latter is manifestly parity-violating. Thus one is led to the
conclusion that unless the conservation of parity is artificially imposed on 
the theory, there is no reason for it to hold generally in a scenario with
torsioned space-time.   

However, the above way of introducing the parity-violating terms under
torsion is of a somewhat {\it ad hoc} nature.
Nor does it provide one with a guideline as to how
to extend this to the gravitational interactions of particles
of different spins, once 
parity ceases to be a symmetry in the pure gravity sector.

As we have already mentioned, our proposal is to seek the very essence of 
parity violation in geometry itself, by introducing it in the
covariant derivative in a curved space-time. That is to say, the most
general connection can be made to include, in addition to the
symmetric and antisymmetric parts, a pseudo-tensorial part as well.
The commutators of covariant derivatives acting on a scalar in such a case
is still non-vanishing, and will depend on both the torsion field and its
pseudo-tensorial extension.
Remembering that the latter should vanish in the limit of zero torsion,
a general form in the minimal extension scheme is

\begin{equation}
\gtil^{\kappa}_{\nu\lambda} = \Gamma^{\kappa}_{\nu\lambda} - 
H^{\kappa}_{\nu\lambda} - 
q(\epsilon_{\nu\lambda}^{\gamma\delta}H_{\gamma\delta}^{\kappa} - 
\epsilon_{\beta\lambda}^{\kappa\alpha}H_{\nu\alpha}^{\beta} + 
\epsilon_{\beta\nu}^{\kappa\alpha}H_{\lambda\alpha}^{\beta}) 
\end{equation}

Here q is a parameter determining the extent of parity violation, depending,
presumably, on the matter distribution. 
It is extremely important to note at this juncture that the covariant 
derivative of the metric ($D_{\mu}g_{\nu\lambda} = 0$, with $D_{\mu}$ 
defined in terms of $\gtil$) automatically 
vanishes so long as H is antisymmetric in all three indices. This means that 
the same condition for metricity as the one in ordinary Einstein-Cartan theory 
also suffices when a pseudo-tensorial
part is included in the connection. However, the relative signs of the three 
additional terms get fixed by
metricity. The same requirement also leads us to the conclusion that all the 
three aforementioned terms have to be coupled through the same charge q.

Next, we calculate the curvature scalar $\tilde{R}$
using the above connection and the metricity condition. After some
algebra, one obtains

\begin{equation}
S = \int \sqrt{-g} \tilde{R} d^{4}x 
\end{equation}

\noindent
with 

\begin{equation}
\tilde{R} = R^{EC} + R^X
\end{equation}

\noindent
where $R^{EC}$ is the ordinary Einstein-Cartan scalar curvature, which
arises from the parity-conserving part of the covariant derivative. The 
artifacts of the pseudo-tensorial connection are found in $R^X$ which is 
given by

\begin{equation}
R^X = R^{pv} + R^{pc}
\end{equation} 

\noindent
with

\begin{equation}
R^{pv} = -6q\epsilon_{\alpha\beta}^{\sigma\nu} H_{\mu}^{\alpha\beta} 
H_{\sigma\nu}^{\mu}
\end{equation} 

\noindent
and

\begin{equation}
R^{pc} = 3q^{2}[2\epsilon_{\lambda\sigma}^{\gamma\delta}
\epsilon_{\omega\nu}^{\sigma\eta}H_{\gamma\delta}^{\nu}
H_{\eta}^{\lambda\omega} + 
\epsilon_{\nu\lambda}^{\rho\eta}
\epsilon_{\alpha\beta}^{\lambda\nu}H_{\sigma}^{\alpha\beta}
H_{\rho\eta}^{\sigma}] 
\end{equation}

The most striking conclusion from above is that the parity violating piece 
$R^{pv}$ is, modulo the multiplicative factor $-6q$,
{\it exactly same}  as what one gets from the
term $\epsilon^{\mu\nu\alpha\beta}R_{\mu\nu\alpha\beta}$ added 
{\it ad hoc} to the Einstein-Hilbert-Cartan action. 
In addition, one would have in the latter case a term proportinal
to the derivative of the H-field, which is absent in our case.
It is obvious, however, that such a derivative term is bound to vanish
whenever torsion can be expressed in terms of the derivative of any
antisymmetric second-rank tensor field, {\it i.e.} 
$H_{\alpha\beta\gamma} = \partial_{[\alpha}B_{\beta\gamma]}$. This is 
precisely what one obtains in superstring theories. Therefore, 
the parity-violating terms obtained in variations of such theories  following
our approach match exactly with those arising from  the term
$\epsilon^{\alpha\beta\mu\nu}R_{\alpha\beta\mu\nu}$ as used in earlier works.
This vindicates our faith in the 
likely origin of parity violation in geometry, as reflected in the 
generalised connection. Moreover, in our aproach,  $R^X$ 
contains terms quadratic in
the Levi-Civita tensor density, which provide extra parity-conserving pieces
over and above those present in the minimal Einstein-Cartan framework.
These terms are, however, proportional to $q^2$, and therefore will be
relatively unimportant when there is only a small amount of parity violation.

It is straightforward now to examine the spin-1 and spin-1/2 sectors.
First, we take up the case of a spin-1 Abelian gauge field A. The 
corresponding field tensor is defined here as 

\begin{equation}
F_{\mu\nu} = D_{\mu}A_{\nu} - D_{\nu}A_{\mu}   
\end{equation}

\noindent
where the covariant derivative is to be written as 
$D_{\mu}A_{\nu} = {\partial}_{\mu}A_{\nu} + 
\gtil^{\beta}_{\mu\nu} A_{\beta}$, which 
includes both the Cartan and pseudo-tensorial extensions. 
Thus the expression for $F_{\mu\nu}$ in our case turns out to be

\begin{eqnarray}
F_{\mu\nu} = \partial_{\mu}A_{\nu} -  \partial_{\nu}A_{\mu}
- 2H^{\beta}_{\mu\nu}A_{\beta}
- 2q\epsilon_{\mu{\nu}}^{\gamma{\delta}}H_{\gamma{\delta}}^{\beta}A_{\beta}
+ 2q\epsilon_{\delta{\nu}}^{\gamma{\beta}}H_{\mu{\beta}}^{\delta}A_{\gamma}
 -2q\epsilon_{\delta{\mu}}^{\gamma{\beta}}H_{\nu{\beta}}^{\delta}A_{\gamma}
\end{eqnarray}

It is obvious that 
the Lagrangian for the gauge field, given by
$-(1/4)F^{\mu\nu} F_{\mu\nu}$, now contains both terms linear and quadratic in 
$\epsilon$. This not only implies parity-violating coupling of the gauge 
field with the background torsion but also brings in an extra parity-conserving
part over and above what is obtained in the the minimal Einstein-Cartan
scenario. Extension of the formalism to a non-Abelian gauge field is
straightforward, with the possibility of additional parity violation 
in the self-coupling terms.

For a spin-1/2 field in a spacetime with torsion, again, the standard 
Dirac Lagrangian needs to be extended with the appropriate covariant 
derivative. It has the general form \cite{matter1,spinor}
  
\begin{eqnarray}
{\cal{L}}_{Dirac} =  {\overline{\psi}}[i \gamma^{\mu}(\partial_{\mu}
-\sigma^{\rho\beta}v_{\rho}^{\nu}g_{\lambda\nu}
\partial_{\mu}v_{\beta}^{\lambda}
- g_{\alpha\delta}\sigma^{ab}v_a^{\beta}v_b^{\delta}
\gtil^{\alpha}_{\mu\beta})]\psi
\end{eqnarray}  

\noindent
where one has introduced the tetrad $v^\mu_a$ to connect the curved space 
with the corresponding tangent space at any point. (The greek indices 
correspond to the curved space, and the Latin indices, to the tangent space.)
Using the full form of $\gtil$ defined in our scenario, 
${\cal{L}}_{Dirac}$ can be expressed as

\begin{eqnarray}
{\cal{L}}_{Dirac} = {\cal{L}}^E + 
{\overline{\psi}}[i\gamma^{\mu}g_{\alpha{\delta}}
\sigma^{\beta{\delta}}H^{\alpha}_{\mu{\beta}}]\psi + {\cal{L}}^{pv}
\end{eqnarray} 

\noindent
the first term being the same as what one would have gotten in
Einstein gravity, and the second one corresponds to the Cartan extension.
The incorporation of a pseudo-tensorial extension results in the 
parity-violating part ${\cal{L}}^{pv}$ which is given by 

\begin{equation}
{\cal{L}}^{pv} = q (det g^{\eta\eta'})[-\overline{\psi}\gamma_{\mu}
\sigma_{\lambda\omega}\gamma_{5}\psi H_{\gamma\rho\delta}  +
g_{\alpha\delta}\overline{\psi}\gamma^{\mu}
\sigma_{\omega\nu}\gamma_{5}\psi H_{\mu\lambda\rho} -
g_{\alpha\delta}\overline{\psi}\gamma_{\mu}
\sigma_{\omega\nu}\gamma_{5}\psi H_{\beta\lambda\rho}] 
\end{equation}
  
\noindent
where $\eta(\eta')$ runs over $\{\delta,\lambda,\omega (\mu,\gamma,\rho)\}$,
$\{\delta,\omega,\nu (\rho,\alpha,\lambda)\}$, and
$\{\beta,\delta,\omega,\nu (\rho,\mu,\alpha,\lambda)\}$ in the first, 
second and third terms.

Note that here one does not get any additional pieces quadratic
in the pseudo-tensorial connecton because of the very structure of the
Dirac Lagrangian. Therefore, unlike in the case of pure gravity and the 
vector field, we do not end up with additional parity-conserving terms
resulting from such a connection.

We conclude by summarising our main points. We have argued here that, since
the very presence of torsion automatically allows  parity-violation in the
Lagrangian for pure gravity, it should be possible to incorporate the latter
in the geometry of space-time itself. We achieve such an end by suitably
extending the covariant derivative with a set of pseudo-tensorial connections,
proportional to the torsion (or more precisely contortion) tensor itself.
We have demostrated that the additional terms are uniquely fixed by the 
condition for preserving metricity. This gives us a consistent
prescription to obtain parity-violating
effects in the Lagrangians for pure gravity, spin-1 and spin-1/2 fields,
as well as new parity-conserving terms in the first two cases. Thus 
parity-violation-- which can be looked upon as the outcome of torsion itself--
can be systematically linked to every sector which is conceivably responsible 
for torsion in space-time.

\noindent
{\bf Acknowledgements}: We thank Ashok Chatterjee, P. Mitra, P. Majumdar, 
A. K. Raychaudhuri and Ashoke Sen for helpful comments. S. S. acknowledges the
hospitality of Mehta Research Institute while this study was
in progress.

\end{document}